\documentclass[3p,twocolumn]{elsarticle}

\usepackage{graphicx}
\usepackage{dcolumn}
\usepackage{pifont}
\usepackage{bm}
\usepackage{multirow}
\usepackage{amsmath}
\usepackage{hyperref}
\usepackage{booktabs}
\usepackage{float}
\usepackage{txfonts}
\usepackage[version=3]{mhchem}

\usepackage[dvips]{color}
\definecolor{darkgreen}{rgb}{0,.6,0}

\journal{J. Mater. Sci.}

\begin{document}
\title{{\it Ab initio} investigation of the adsorption of zoledronic acid molecule on hydroxyapatite (001) surface: an atomistic insight of bone protection}

\author[kimuniv-m]{Mun-Hyok Ri}
\author[kimuniv-m]{Chol-Jun Yu\corref{cor}}
\ead{ryongnam14@yahoo.com}
\author[kimuniv-c]{Yong-Man Jang}
\author[kimuniv-c]{Song-Un Kim}

\cortext[cor]{Corresponding author}

\address[kimuniv-m]{Department of Computational Materials Design (CMD), Faculty of Materials Science, and}
\address[kimuniv-c]{Department of Organic Chemistry, Faculty of Chemistry, Kim Il Sung University, Ryongnam-Dong, Taesong-District, Pyongyang, DPR Korea}

\begin{abstract}
We report a computational study of the adsorption of zoledronic acid molecule on hydroxyapatite (001) surface within {\it ab initio} density functional theory. The systematic study has been performed, from hydroxyapatite bulk and surface, and zoledronic acid molecule to the adsorption of the molecule on the surface. The optimized bond lengths and bond angles were obtained and analyzed, giving an evidence of structural similarity between subjects under study. The formation energies of hydroxyapatite (001) surfaces with two kinds of terminations were computed as about 1.2 and 1.5 J/m$^2$ with detailed atomistic structural information. We determined the adsorption energies of zoledronic acid molecule on the surfaces, which are -260 kJ/mol at 0.25 ML and -400 kJ/mol at 0.5 ML. An atomistic insight of strong binding affinity of zoledronic acid to the hydroxyapatite surface was given and discussed.
\end{abstract}


\maketitle

\section{\label{sec:intro}Introduction}
Bisphosphonates (BPs) are widely used as the most powerful drug for protecting bone and treating bone disease such as osteoporosis and other metabolic bone disorders~\cite{Ebetino,Bartl,Coxon,Russell08,Russell11}. From the 1960s onward, a great number of extensive studies on BPs with respect to the development and clinical treatment have been carried out, but the research field on BPs is still very active, being continued to evolve towards a better understanding of treatment mechanisms and a finding of more potent BP on the basis of it. In this context, zoledronic acid (or zoledronate, ZOD), whose chemical name is [1-hydroxy-2-(1H-imidazol-1) ethylidene] bisphosphonic acid or 2-(imidazol-1-yl)-1-hydroxy-ethane-1,1-diphosphonic acid, was developed as a third-generation BP and was approved for the oral treatment of bone disease in 2012 in DPR Korea after the success of its synthesis in our own way by two of the authors (Yong-Man Jang and Song-Un Kim).

BPs are characterized by two phosphonate groups, central carbon atom in between, and two side groups R1 and R2, while human bone is composed of complex array of hydroxyapatite (HAP, \ce{Ca5(PO4)3OH}) crystallites with nano sizes ranging from 30 to 200 nm embedded within the collagen matrix~\cite{Hassenkam,Henning,Elliot}. It is well established that the function of BPs to inhibit bone resorption is originated with their ability of binding strongly to bone mineral, that is, HAP crystal, and stiff resistance to chemical and enzymatic hydrolysis~\cite{Russell08}. In more detailed explanation, the P-C-P backbone of BPs made from two phosphonate groups and carbon atom has a high binding affinity for the compositions of HAP -- tetrahedral PO$_4$ groups, OH groups and Ca ions -- with an additional contribution from the R1 side group (in most cases including ZOD, it is -OH)~\cite{Russell08,Russell}. Moreover, the P-C-P group of BPs is considerably more resistant to the dissolution of HAP crystal than the P-O-P group of pyrophosphate, of which BPs are stable structural analogues~\cite{Kontecka,Fleich,Rogers,Roelofs,Dunford}.

Varying the another side group R2 of BPs can result in differences in antiresorptive potency with several orders of magnitude. It was observed that more potent BPs posses a primary, secondary or tertiary nitrogen atom in the R2 side chain~\cite{Nancollas,Lawson}. At present, the most potent antiresorptive BPs include a heterocyclic R2 side chain containing a nitrogen atom like risedronate and zoledronate~\cite{Green}. The structure-activity relationship analysis showed that the strong binding affinity of zoledronate for HAP is related to its 3D shape and atomic orientation, indicating an important role of 3D shape of nitrogen-containing BP and the orientation of its nitrogen in binding affinity for HAP~\cite{Russell08}.

Atomistic modeling and simulations are a powerful tool to describe the structural characteristics of BPs and bone mineral HAP and the interaction between them at atomic scale, as proved in materials science and molecular science through a vast number of applications. For instance, molecular dynamics (MD) simulations based on the well-constructed classical force field have provided the structural information and energetics of bone mineral, HAP crystal and its surfaces~\cite{Bhowmik,Zhu, Matsunaga, Barrios}. Bhowmik et al.~\cite{Bhowmik} obtained the structural parameters of monoclinic HAP crystal and its surface energetics using the consistent valence force field (CVFF), presenting a valuable description of the interaction between polyacrylic acid and HAP. The surface energetics of HAP crystalline surfaces using {\it ab initio} density functional theory (DFT) calculations within the generalized gradient approximation (GGA) for the exchange-correlation functional has been studied by Zhu and Wu~\cite{Zhu}, testing the effects of slab thickness, vacuum width between slabs and surface relaxation on surface energy. Barrios~\cite{Barrios} has investigated the interaction between collagen protein and HAP surface by using a combination of computational techniques, DFT and classical MD methods. Duarte et al.~\cite{Duarte} performed molecular mechanics simulations for molecular structures of 18 novel hydroxyl- and amino-bisphosphonates to examine the interaction between BPs and hydroxyapatite and to extract relating structural characteristics of BPs and their affinities for the mineral, which are in agreement with {\it in vitro} and {\it in vivo} studies for some of the studied BPs. To the best our knowledge, however, investigation on surface adsorption of zoledronate on HAP surface with its detailed atomistic structure based on quantum mechanics is still scarce, in spite of such extensive theoretical studies of BPs and HAP surface, and we believe that {\it ab initio} study on this phenomenon should definitely contribute to a better understanding of the interaction of zoledronate with bone mineral at atomic and electronic scale.

In this paper, we carry out systematic {\it ab initio} DFT calculations for zoledronate molecule, hydroxyapatite bulk and surface, and surface adsorption of zoledronate molecule on hydroxyapatite surface. Our special focus is placed on the adsorption of zoledronate molecule on hydroxyapatite surface, providing the adsorption energy and atomistic structures of surface complexes, and an insight how charge transferring is occurred in the event of adsorption. This is aimed to get a reliable insight for bone protection effect of zoledronate at electronic scale. In the following, we first describe a computational method in Sec.~\ref{sec:method}, present the results for structural parameters and electronic properties of hydroxyapatite bulk and surface, zoledronate molecule, and for binding of zoledronate to hydroxyapatite surface in Sec.~\ref{sec:result}, and finally give our conclusions in Sec.~\ref{sec:con}.

\section{\label{sec:method}Computational Method}
For the DFT calculations in this work, we have employed \textsc{SIESTA} code\cite{SIESTA} which solves numerically Kohn-Sham equation within DFT~\cite{Hoh1964,Koh1965} using a localized numerical basis set, namely pseudo atomic orbitals (PAO), and pseudopotentials for describing the interaction between ionic core (nucleus plus core electrons) and valence electrons. The BLYP GGA functional (the Becke exchange functional~\cite{Becke} in conjunction with the Lee-Yang-Parr correlation functional~\cite{LYP}) was used for exchange--correlation interaction between electrons. For all the atoms, Troullier-Martins~\cite{TMpseudo} type norm-conserving pseudopotentials were generated within local density approximation (LDA)~\cite{PZlda}, and checked carefully. The basis sets used in this work were the DZP type (double $\zeta$ plus polarization). The mesh size of grid, which is controlled by energy cutoff to set the wavelength of the shortest plane wave represented on the grid, has taken a value of 200 Ry. Non-fixed atoms were allowed to relax until the forces converge less than 0.02 eV/\AA.

We first determine the crystal lattice parameters of bulk HAP by performing structural relaxation with atomic coordinates optimization using the conjugate gradient scheme and appropriate Monkhorst-Pack k-points. Through a comparison of the lattice constants with the experimental values, we have a confidence of the above mentioned selection for calculation parameters. Structural optimization for isolated ZOD molecule is performed as well, and this for several conformations to select the most stable one, of which total energy is the lowest among studied conformations.

We then select the interesting surface index as (001) on the basis of experimental evidence that this face provides binding sites for many ionic species~\cite{Barrios09,Wierzbicki}. In order to simulate the surface within a three-dimensional simulation code, we use two-dimensional periodic slabs, which have a thickness of atomic layers of 2 crystal unit cells plus 30~\AA~vacuum layer, allowing the atoms in one half of a crystal unit cell at each side of the slab to relax. These structural parameters in the supercell can give well-converged result: increasing the thickness of slab to 3 and 4 crystal unit cells and the vacuum width up to 50~\AA~result in variations of surface energy smaller than $\pm$0.05 J/m$^2$. To check the stability of the surface, the surface formation energies are calculated approximately as the difference between the total energies of the slab and the corresponding bulk crystal:
\begin{equation}
\gamma=\left(E_\text{slab}-\frac{N_\text{slab}}{N_\text{bulk}}E_\text{bulk}\right)/2A, \label{eq:gamma}
\end{equation}
where $N_\text{slab}$ and $N_\text{bulk}$ are the numbers of atoms in the surface slab and in the bulk unit cell, $A$ is the slab surface area, and $E_\text{slab}$ and $E_\text{bulk}$ are the total energies of the slab and bulk structures, respectively~\cite{yucj}. The sign of surface formation energy is a test of surface stability: positive (negative) means energy should be provided (released) in order to create a surface. We note that there might be several possible cutting planes for a particular Miller index while guaranteeing neutrality of the surface charge perpendicular to the surface\cite{tasker}.

Final step of the work is to simulate the adsorption of ZOD molecule on the relevant relaxed HAP (001) surfaces. In order to have the rough, initial surface adsorption structure, we use classical molecular mechanics (MM) method. In this MM simulation, General Utility Lattice Program (GULP)~\cite{GULP} is utilized. For describing the interaction between atoms, we use the Dreiding force field, where the potential energy is described as the sum of the contributions resulting from the bonded interactions (bond stretching, bond bending, and torsions) and from the non-bonded interactions (e.g., electrostatic interaction, and van der Waals interaction)~\cite{Leach}. After making a guess for the initial state of the adsorption complex, we also carry out atomic relaxation for the surface complex -- ZOD molecule and surface atoms in the slab -- to obtain a stable final state. Then, the adsorption energy can be calculated as follows:
\begin{equation}
E_\text{ads}=\frac{1}{N_\text{mol}}\left[E_\text{mol-slab}-(E_\text{slab}+N_\text{mol}E_\text{mol})\right],
\label{eq:Ehydro}
\end{equation}
where $N_\text{mol}$ is the number of adsorbed molecules, and $E_\text{mol-slab}$ and $E_\text{mol}$ are the total energies of the adsorbed surface and of the isolated molecule, respectively. The adsorption energy can be either negative or positive: negative (positive) means energy should be released (provided) during the adsorption, indicating that the adsorption is (not) spontaneous exothermic (endothermic) process.

\section{\label{sec:result}Results and discussion}
\subsection{\label{subsec:bulk}Bulk hydroxyapatite}
The crystal structure of hydroxyapatite is hexagonal with space group $P6_3/m$, which contains a formula unit \ce{Ca10(PO4)6(OH)2}. In fact, there must be four hydroxyl groups in the unit cell with the $P6_3/m$ space group, each oxygen atom of hydroxyl group with 1/2 occupancy~\cite{Barrios}. To make the DFT calculations enable, therefore, we have made a model with full occupancies for the hydroxyl groups, by assigning alternate 0 and 1 occupancies to these hydroxyl groups, which results in the change of the space group from $P6_3/m$ to $P6_3$. However, the modified model has a net electric polarization contrary to the experiment which shows zero polarization, because all the hydroxyl groups in the model are oriented in the same direction. To mimic the real structure where exists disorder in the relative orientation of the parallel OH channels, so that electric polarizations are compensated each other, we have created a supercell containing two unit cells in the $[100]$ direction, and assigning opposite orientations to the two OH channels in the supercell. The crystal structure with antiparallel hydroxyl groups in a double unit cell compared to the original hexagonal structure is monoclinic with $P2_1$ space group. We have used this structure in this work, thus allowing a direct comparison of the simulation results with experimentally determined surfaces.

\begin{figure}[!ht]
\includegraphics[clip=true,scale=0.28]{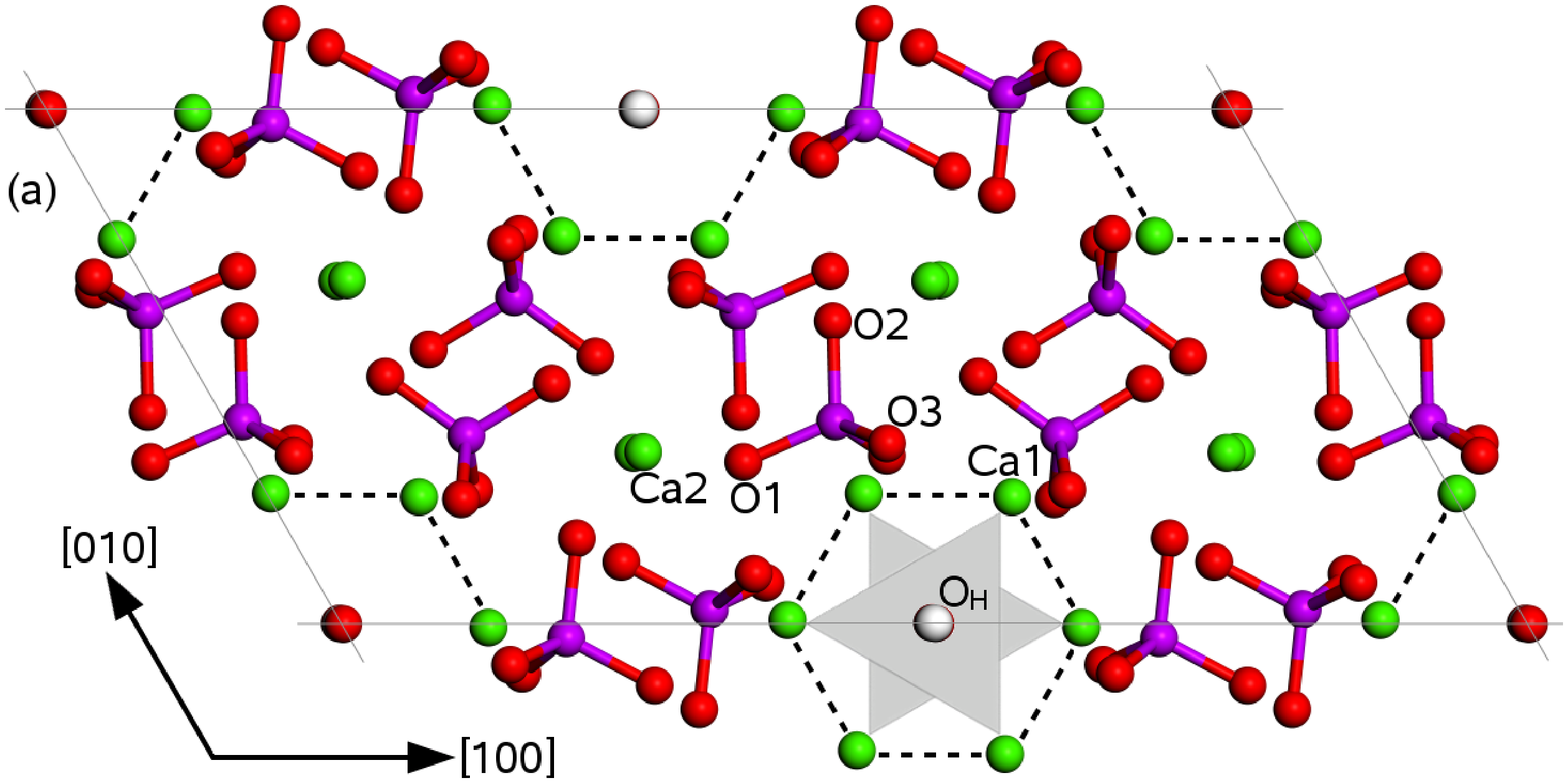} \\ \vspace{1pt} \hspace{16pt}
\includegraphics[clip=true,scale=0.265]{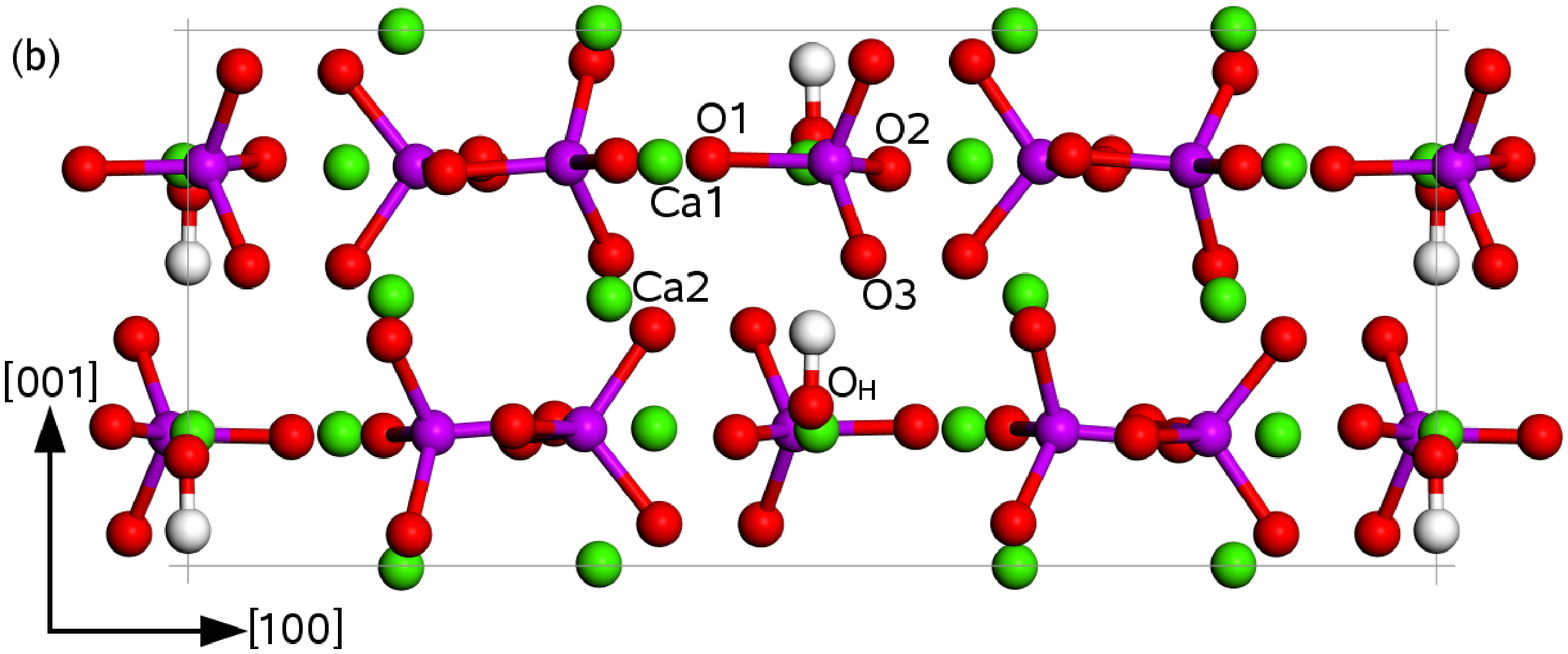}
\caption{\label{fig:bulk}(Color online) Top view (a) and side view (b) of supercell of bulk hydroxyapatite crystal with monoclinic $P2_1$ space group, which is doubled the unit cell in $[100]$ direction so as to compensate the electric polarization by assigning opposite orientations to the two OH channels in hexagon surrounded by triangle of Ca1 ions. (a) top view and (b) side view. (Ca: Green, P: purple, O: red, and H: white)}
\end{figure}
The full optimization of the structure, allowing the cell shape and volume, and ionic positions to relax, was performed. We have checked the convergence with respect to the special k-points; increasing the k-points set from $(1\times2\times3)$ to $(2\times4\times6)$ leads to the change between the total energies as about 5 meV, thus indicating the safe use of $(1\times2\times3)$ set. Figure~\ref{fig:bulk} depicts the fully optimized supercell of bulk HAP in this work. The calculated structural parameters and chemical bonding properties of bulk HAP are given in Table~\ref{tab:optlattice}. The lattice constants ($a$=9.348, $b$=9.352, and $c$=6.955 \AA) of the fully optimized structure are in good agreement with the experimental values ($a$=$b$=9.43, $c$=6.891 \AA)~\cite{Kim, Posner} (less than 1.0\% error) as well as with the earlier DFT results~\cite{Leeuw02, Ellis, Barrios}. Note that $a$ is the half of lattice constant of the supercell.
\begin{table}[!ht]
\begin{center}
\small 
\caption{\label{tab:optlattice}Structural parameters of bulk hydroxyapatite crystal structure with monoclinic $P2_1$ space group, compared with experimental results.}
\begin{tabular}{lcc}
\toprule
 &\multicolumn{2}{c}{Lattice parameters} \\
 \cline{2-3}
 & This work & Experiment$^a$ \\
 \cline{2-3}
 $a, b, c$ (\AA) & 9.348, 9.352, 6.955 & 9.430, 9.430, 6.891 \\
 $\alpha, \beta, \gamma$ ($^\circ$) & 90, 90, 119.86 & 90, 90, 120 \\
\hline
 & \multicolumn{2}{c}{Average of bond lengths (\AA)} \\
 \cline{2-3}
P$-$O              & 1.578 & 1.540 \\
Ca1$-$O            & 2.423 & 2.478 \\
Ca1$-$O$_\text{H}$ & 2.339 & 2.354\\
Ca2$-$O            & 2.553 & 2.556 \\
\hline
 & \multicolumn{2}{c}{Average of bond angles ($^\circ$)} \\
 \cline{2-3}
O$-$P$-$O & 109.44 & 109.45 \\
O$-$Ca1$-$O & 98.62 &  \\
O$-$Ca2$-$O & 99.40 &  \\
\bottomrule
\end{tabular} \\
{\raggedright
$^a$ Ref.~\cite{Kim}\\
}
\end{center}
\end{table}
\normalsize

As shown in Figure~\ref{fig:bulk}, there are two different Ca sites (denoted as Ca1 and Ca2), four oxygen sites (O1, O2, and O3 of phosphate tetrahedron, and O$_\text{H}$ of hydroxyl group), one P site, and one H site. The Ca1 atom is coordinated to seven oxygen ions, which are six oxygens of different five \ce{PO4} groups and one oxygen of OH group, while the Ca2 atom has a ninefold coordination with oxygen ions situated in six different \ce{PO4} tetrahedra. The three Ca1 atoms also form a triangle at the same plane normal to the $c$ axis, whose center is occupied by an OH group, and the two triangles form a hexagonal screw configuration. The calculations yield average bond lengths as 1.578~\AA~for P-O bond, 2.423~\AA~for Ca1$-$O, 2.339~\AA~for Ca1$-$O$_\text{H}$, and 2.553~\AA~for Ca2$-$O, compared to the corresponding experimental values of 1.54, 2.478, 2.354, and 2.556~\AA, respectively. The similar agreement is obtained in the bond angles, as shown in Table~\ref{tab:optlattice}. These comparisons confirm that our computational parameters are reasonably satisfactory for a good assessment of surface calculations.

\subsection{\label{subsec:surf}Hydroxyapatite (001) Surface}
In this work, we have selected the Miller indices of HAP surface as (001), since there exists experimental evidence that the (001) surface is the most stable among several surfaces with low indices and provides binding sites for many ionic species~\cite{Leeuw04}. As we use the supercell doubled in the [100] direction as a unit cell for bulk HAP, the (001) surface unit cell is an oblique (2$\times$1) surface cell and there might be two kinds of terminations; (1) 2Ca2$-($6PO$_4\cdot$6Ca1$\cdot$2OH)$-$2Ca2, denoted as Type (I), and (2) (3PO$_4\cdot$3Ca1$\cdot$OH)$-$4Ca2$-($3PO$_4\cdot$3Ca1$\cdot$OH), denoted as Type (II). They are repeated in the [001] direction, assigned to building layer, and belong to the Tasker (II)-type surface~\cite{tasker} with no electric dipole moment perpendicular to the surface.

\begin{figure*}[!ht]
\begin{center}
\includegraphics[clip=true,scale=0.45]{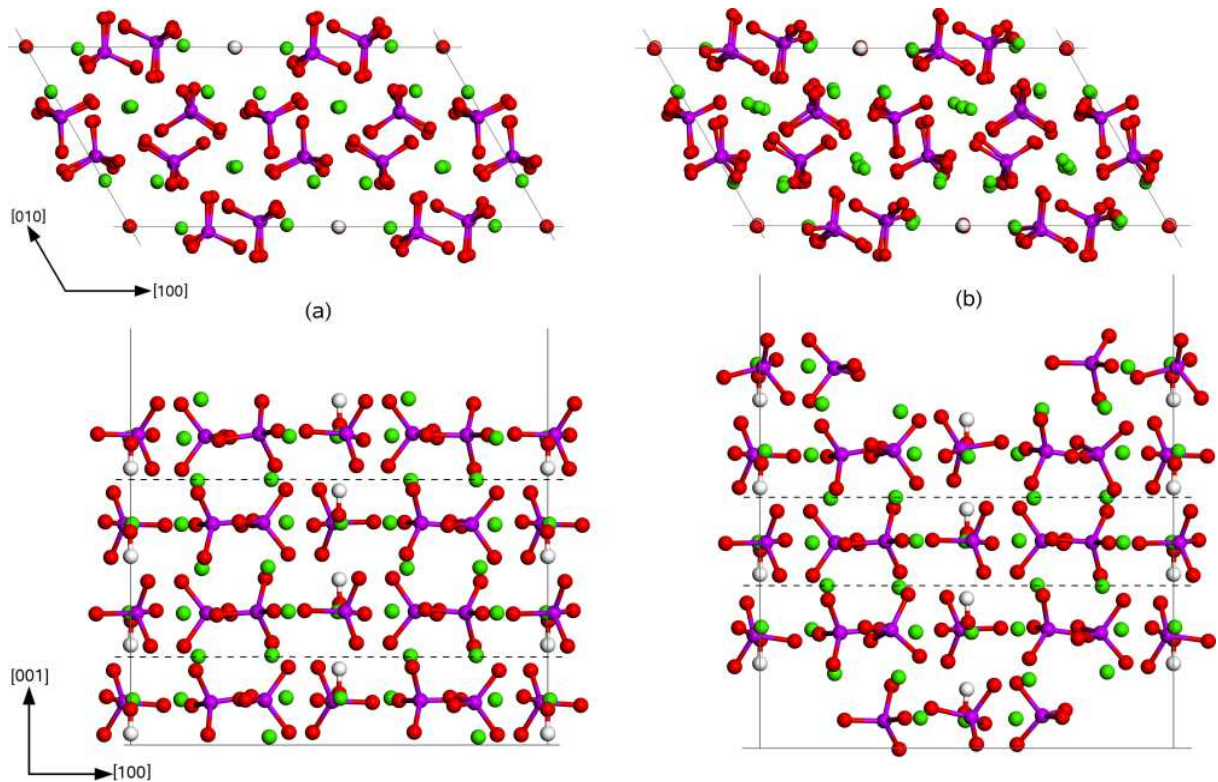}
\caption{\label{fig:surf}(Color online) Top view (upper panel) and side view (lower panel) of optimized structure of hydroxyapatite (001) surface with (2$\times$1) surface unit cell, vacuum thickness of 30 \AA, 4 building layers, and termination 2Ca2$-($6PO$_4\cdot$6Ca1$\cdot$2OH)$-$2Ca2 (a) and termination (3PO$_4\cdot$3Ca1$\cdot$OH)$-$4Ca2$-($3PO$_4\cdot$3Ca1$\cdot$OH) (b). The top and bottom layers are allowed to relax, while the layers between dashed lines are fixed at their crystalline positions. (Ca: green, P: purple, O: red, H: white)}
\end{center}
\end{figure*}
\begin{table}[!ht]
\begin{center}
\small 
\caption{\label{tab:surfene}Surface energies (J/m$^2$) according to vacuum thickness and building layers of hydroxyapatite (001) surfaces with (2$\times$1) surface unit cell. The values in parenthesis represent the number of atoms.}
\begin{tabular}{lllll}
\toprule
\multicolumn{5}{c}{Type (I); 2Ca2$-($6PO$_4\cdot$6Ca1$\cdot$2OH)$-$2Ca2} \\
\cline{2-5}
                              & Size & Unrelaxed & Relaxed & Ref.$^a$ \\ 
\cline{2-5}
\multirow{4}{*}{Vacuum (\AA)} & 20 & 1.460 & 1.208 &  \\ 
                              & 30 & 1.461 & 1.205 &  \\ 
                              & 40 & 1.448 & 1.197 &  \\ 
                              & 50 & 1.449 & 1.205 &  \\ 
\cline{2-5}
\multirow{3}{*}{Layer}        & 4 (176) & 1.461 & 1.205 &  \\ 
                              & 6 (264) & 1.474 & 1.218 &  \\ 
                              & 8 (352) & 1.496 &       &  \\ 
\hline
\multicolumn{5}{c}{Type (II); (3PO$_4\cdot$3Ca1$\cdot$OH)$-$4Ca2$-($3PO$_4\cdot$3Ca1$\cdot$OH)} \\
\cline{2-5}
\multirow{3}{*}{Layer}        & 4 (176) & 2.084 & 1.514 & 1.01 \\ 
                              & 6 (264) & 2.057 & 1.506 &  \\ 
                              & 8 (352) & 2.078 &  &  \\ 
\bottomrule
\end{tabular} \\
{\raggedright $^a$ Ref.~\cite{Barrios} \\}
\end{center}
\end{table}
\normalsize
The HAP (001) surface has been modelled using three-dimensional periodic supercell (slab) with two equivalent surfaces at bottom and top side. To ensure no interaction between bottom surface of the above image slab and top surface of the present slab across the vacuum region, the vacuum region must be wide enough. And the atomic layer, which is consisted of surface layer allowed to relax and crystal layer fixed at its crystalline position, should be thick enough so that the two surfaces of each slab do not interact through the crystal layer. To check the convergence according to the vacuum region, we tested 20, 30, 40, and 50 \AA~thickness, and confirmed that there is no distinct change between 20 and 50 \AA~thick vacuums (Table~\ref{tab:surfene}). Therefore, the vacuum region of 30 \AA~thickness will be used in the following calculations. The thickness of the slab is usually expressed in terms of a number of building layers, where one layer contains 44 atoms. The four layers (two bulk unit cells, 176 atoms), six layers (three unit cells, 264 atoms), and eight layers (four unit cells, 352 atoms) were tested with the vacuum width of 30 \AA, and the change of surface energies between 4 layers slab and 8 layers slab is only 0.05 J/m$^2$. Thus we will use a slab with 4 building layers in the study of surface adsorption.

As listed in Table~\ref{tab:surfene}, the surface energy of Type (II) surface is slightly higher than that of Type (I) surface, indicating the Type (I) terminated surface is more favorable than the Type (II) terminated surface. We see that the surface relaxation in the Type (I) surface is not really much compared to the Type (II) surface, since the difference between unrelaxed and relaxed surface energies in the former case is not remarkable contrary to the latter case. Since there is no data available in the literature for Type (I), we only compared the calculated Type (II) surface energy with the previous SIESTA work~\cite{Barrios}.

Figure~\ref{fig:surf} shows the fully relaxed atomistic structure of the HAP (001) surfaces modelled by a slab with vacuum thickness of 30 \AA~and four building layers. It is observed that the coordination number (CN) of Ca1 atom changes from 7 in bulk to 6, and CN of Ca2 from 9 to 6 in the Type (I) termination, while in the Type (II) termination the CN of Ca1 atom changes from 7 to 6, and Ca2 atom from 9 to 5, and 6. The coordination environments around the surface oxygen atoms (O1, O2, O3, and O$_\text{H}$) are also changed from the their bulk environment, while P atoms are still fully surrounded by four oxygen atoms like in bulk.

Table~\ref{tab:surfstruct} shows the bond lengths of cation-oxygen and bond angles of oxygen-cation-oxygen at the top surface layer.
\begin{table}[!ht]
\begin{center}
\small 
\caption{\label{tab:surfstruct}Bond lengths of cation-oxygen and bond angles of oxygen-cation-oxygen at hydroxyapatite (001) surface with coordination numbers (CN) of cations.}
\begin{tabular}{lllll}
\toprule
\multicolumn{5}{c}{Type (I); 2Ca2$-($6PO$_4\cdot$6Ca1$\cdot$2OH)$-$2Ca2} \\
\cline{2-5}
    &       & \multicolumn{2}{c}{Angle ($^\circ$)} &  \\ 
\cline{3-4}
    & Length (\AA) &        Range      & Average & CN \\
\cline{2-5}
P   & 1.583 & 103.47$\sim$116.23 & 109.39 & 4 \\
Ca1 & 2.352 & 64.54$\sim$155.26  & 101.50 & 6 \\
Ca2 & 2.434 & 62.45$\sim$137.08  & 94.01  & 6 \\
\hline
\multicolumn{5}{c}{Type (II); (3PO$_4\cdot$3Ca1$\cdot$OH)$-$4Ca2$-($3PO$_4\cdot$3Ca1$\cdot$OH)} \\
\cline{2-5}
P   & 1.585 & 102.07$\sim$117.07 & 109.22 & 4 \\
Ca1 & 2.420 & 60.97$\sim$159.10  & 99.72  & 6 \\
Ca1 & 2.343 & 64.73$\sim$160.44  & 102.36 & 6 \\
Ca2 & 2.325 & 66.52$\sim$147.60  & 101.91 & 5 \\
Ca2 & 2.414 & 60.91$\sim$170.66  & 98.61  & 6 \\
Ca2 & 2.380 & 61.52$\sim$154.20  & 99.52  & 6 \\
\bottomrule
\end{tabular}
\end{center}
\end{table}
\normalsize
In the case of Type (II) surface, we consider the two kinds of Ca1 and three kinds of Ca2 surface atoms. In both cases, the averages (1.583, 1.585 \AA) of bond lengths of P$-$O are a bit expanded compared with the bulk (1.578 \AA), and the averages of bond angles get smaller than in the bulk. The bond lengths of Ca2$-$O are clearly contracted at the surface; 2.434 in Type (I), and 2.325, 2.380, and 2.414 \AA~in Type (II), which are all smaller than the bulk value 2.553 \AA. From these observations, it can be concluded that the undercoordinated surface Ca atoms and O atoms can be favorable adsorption sites.

\subsection{\label{subsec:zol}Zoledronic acid molecule}
\begin{figure}[!ht]
\begin{center}
\includegraphics[clip=true,scale=0.65]{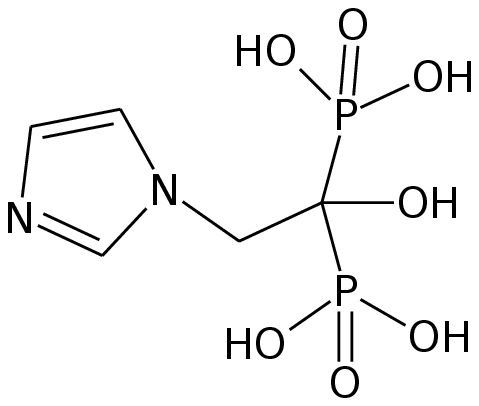} \\
\includegraphics[clip=true,scale=0.40]{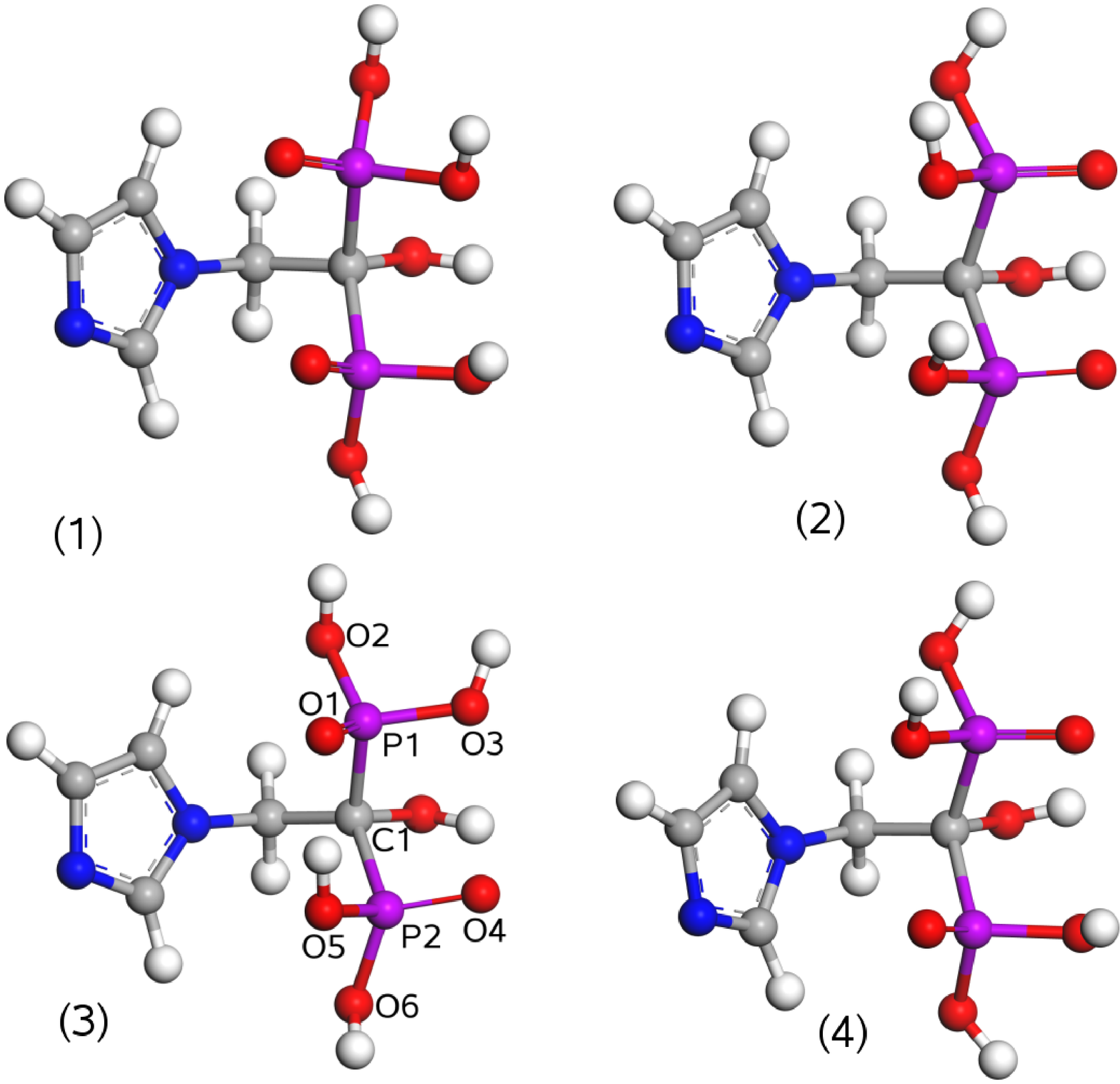} \\
\includegraphics[clip=true,scale=0.40]{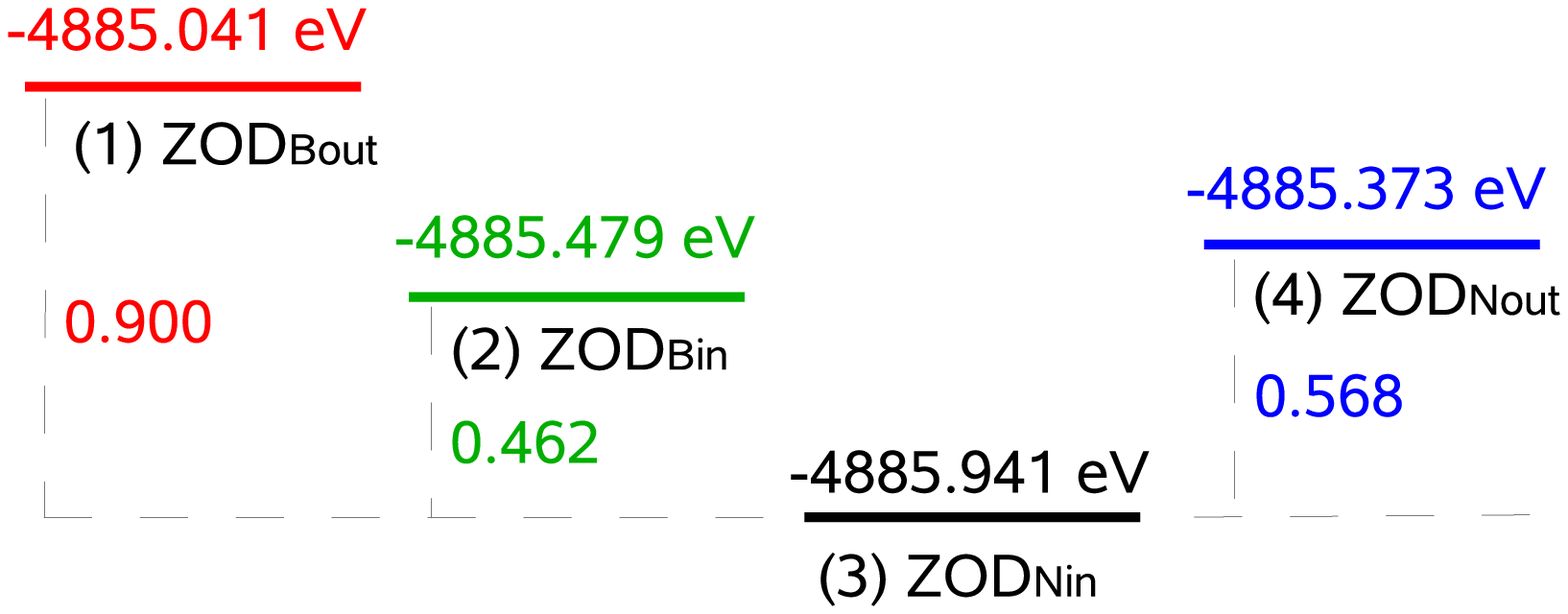} 
\caption{\label{fig:zol}(Color online) Optimized structures of zoledronic acid with four different conformations, (1) ZOD$_\text{Bout}$, (2) ZOD$_\text{Bin}$, (3) ZOD$_\text{Nin}$, and (4) ZOD$_\text{Nout}$. (C: grey, P: purple, N: blue, O: red, H: white)}
\end{center}
\end{figure}

To simulate an isolated molecule, we have used a cubic supercell with lattice constants of $a=b=c=50$~\AA, which has been proved to be enough to prevent the artificial interaction between adjacent molecules. There might be four different conformations, as presented in Figure~\ref{fig:zol}. We can make a distinction between the conformations according to the directions of two pairs of OH groups attached to the two P atoms; (1) ZOD$_\text{Bout}$ for the case where the directions of both OH pairs are outward against the nitrogen heterocyclic ring, (2) ZOD$_\text{Bin}$ for the case of inward directions of both pairs, (3) ZOD$_\text{Nin}$ for the case of inward direction of one pair in the same side of nitrogen atom, and (4) ZOD$_\text{Nout}$ for the case of outward direction of one pair in the N side. After the total energy minimizations to get the optimized structure of molecule, we have compared the total energies of four conformations. The calculations tell us the most stable structure is just the third case, ZOD$_\text{Nin}$, which will be used in the following adsorption study.

The optimized bond lengths and bond angles related with P atoms in ZOD$_\text{Nin}$ conformation are listed in Table~\ref{tab:zol}. The averages of P$-$O bond lengths are 1.597 \AA~in P1 side and 1.586 \AA~in P2 side, which are similar to those of bulk HAP (1.578 \AA) and of HAP (001) surface (1.583 \AA). The averages of O$-$P$-$O bond angles (112.11 and 111.42$^\circ$) are also close to those of bulk HAP (109.44$^\circ$) and HAP surface (109.39$^\circ$). It is found that those values of P2 side are slightly closer to the bulk and surface values than P1 side, so that P2 side are more favorable to binding with HAP due to the closer structural similarity.
\begin{table}[!ht]
\begin{center}
\small
\caption{\label{tab:zol}Optimized bond lengths and angles related with phosphorus atoms in zoledronic acid conformation, ZOD$_\text{Nin}$. Ave. means average.}
\begin{tabular}{ll|ll}
\toprule
\multicolumn{4}{c}{P$_1$ related} \\
\hline
\multicolumn{2}{c|}{Bond length (\AA)} & \multicolumn{2}{c}{Bond angle ($^\circ$)} \\
\hline
P1$=$O1 & 1.495 & O1$=$P1$-$O2 & 117.26 \\
P1$-$O2 & 1.654 & O1$=$P1$-$O3 & 115.10 \\
P1$-$O3 & 1.641 & O1$=$P1$-$C1 & 115.22 \\
P1$-$C1 & 1.902 & O2$-$P1$-$O3 & 103.97 \\
        &       & O2$-$P1$-$C1 & 102.96 \\
        &       & O3$-$P1$-$C1 & 100.12 \\
        &  Ave. & O$-$P1$-$O   & 112.11 \\
        &  Ave. & O$-$P1$-$C1  & 106.10 \\
\hline
\multicolumn{4}{c}{P$_2$ related} \\
\hline
P2$=$O4 & 1.495 & O4$=$P2$-$O5 & 120.01 \\
P2$-$O5 & 1.622 & O4$=$P2$-$O6 & 117.35 \\
P2$-$O6 & 1.641 & O4$=$P2$-$C1 &  99.93 \\
P2$-$C1 & 1.934 & O5$-$P2$-$O6 &  96.91 \\
        &       & O5$-$P2$-$C1 & 112.73 \\
        &       & O6$-$P2$-$C1 & 110.40 \\
        &  Ave. & O$-$P2$-$O   & 111.42 \\
        &  Ave. & O$-$P2$-$C1  & 107.69 \\
\hline
        &       & P1$-$C1$-$P2 &  98.65 \\
\bottomrule
\end{tabular}
\end{center}
\end{table}
\normalsize

\begin{figure}[!ht]
\begin{center}
\includegraphics[clip=true,scale=0.42]{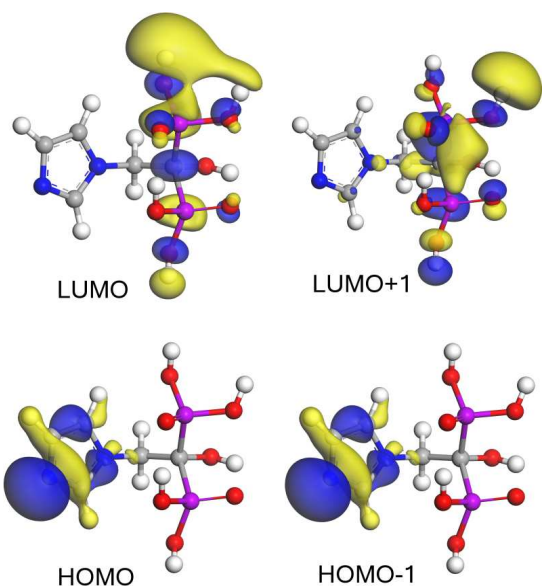}
\caption{\label{fig:homo}(Color online) Molecular orbitals of zoledronic acid conformation, ZOD$_\text{Nin}$.}
\end{center}
\end{figure}
Figure~\ref{fig:homo} shows the computed isodensity surfaces for the highest occupied molecular orbitals (HOMO) and lowest unoccupied molecular orbitals (LUMO) of ZOD (conformation ZOD$_\text{Nin}$). While the HOMO and HOMO-1 are localized on the heterocyclic ring containing nitrogen atom, LUMO and LUMO+1 are on the phosphonate groups. This leads to the intramolecular charge separation upon excitation. The change in electronic distribution between HOMO and LUMO indicates that two phosphonate groups can play a role of electron acceptor in the chemical reaction, while the heterocyclic ring containing nitrogen atom could be a donor.

\subsection{\label{subsec:ads}Adsorption of zoledronic acid on hydroxyapatite (001) surface}
To begin with adsorption, we have utilized GULP to obtain rough structure of ZOD absorbed on HAP (001) surface, where Dreiding forcefield was adopted. Simulated annealing was performed, increasing the temperature from 300 K to 10000 K and subsequently decreasing back to 300 K with the temperature interval of 50 K. Using the obtained rough structure as the starting structure, we then performed atomic relaxation with SIESTA code to get the final optimized structure of zoledronic acid adsorbed HAP (001) surface.

\begin{figure*}[!ht]
\begin{center}
\includegraphics[clip=true,scale=0.45]{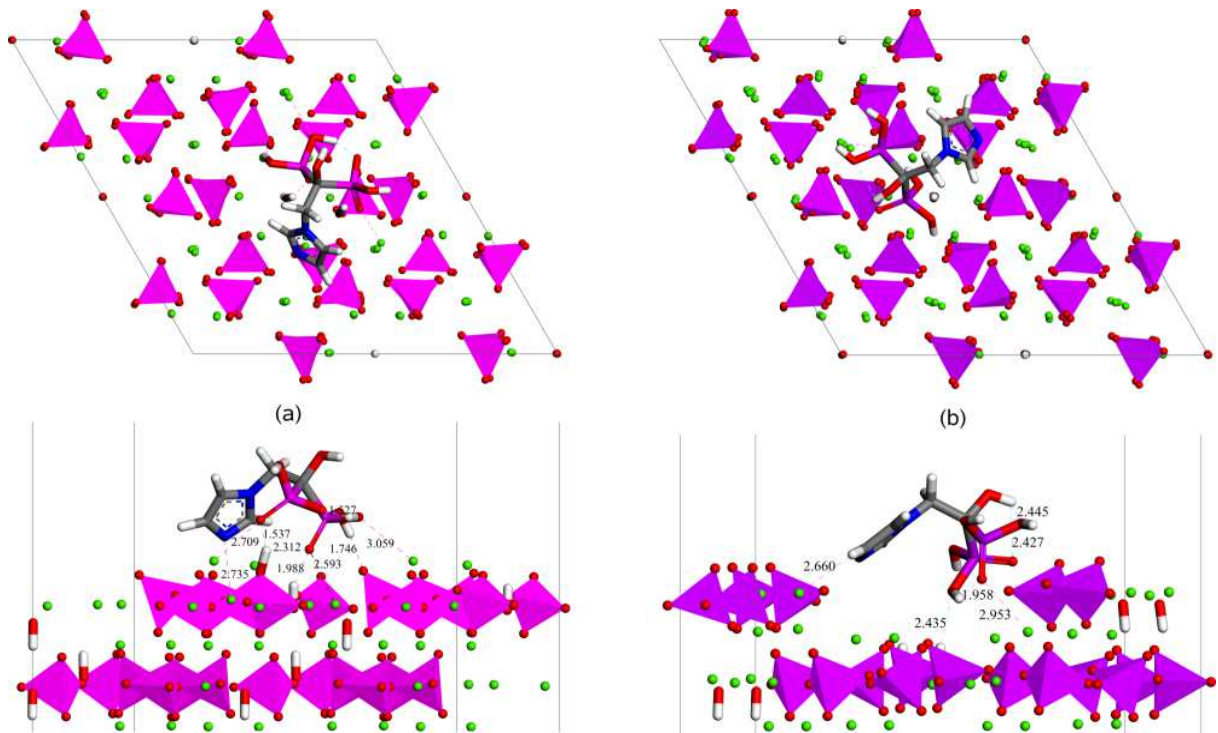}
\caption{\label{fig:ads}(Color online) Top view (upper panel) and side view (lower panel) of optimized atomistic structure of zoledronic acid adsorption complexes on hydroxyapatite (001) surfaces with Type (I) (a) and Type (II) (b) terminations at 0.25 ML coverage. (Ca: green, P: purple, O: red, N: blue, H: white)}
\end{center}
\end{figure*}

In this work we have tried to simulate the adsorption of ZOD molecule at 0.5 ML (one molecule on (2$\times$1) surface cell) and at 0.25 ML coverages (one molecule on (2$\times$2) surface cell). As mentioned above, there are two possible terminations in the HAP (001) surface, and therefore four kinds of simulation tasks were carried out. We have used the slab model with 4 building layers and vacuum thickness of 30 \AA, which guarantee to give a reliable adsorption energy. To make a systematic error small, the pristine surface energies for (2$\times$2) surface slabs were also calculated.
\begin{table}[!ht]
\begin{center}
\small
\caption{\label{tab:ads-ene}Calculated adsorption energies (kJ/mol) of zoledronic acid on hydroxyapatite (001) surface.}
\begin{tabular}{llll}
\toprule
  & \multicolumn{2}{c}{Coverage} \\
\cline{2-3}
  & 0.5 ML (2$\times$1) & 0.25 ML (2$\times$2) \\
\hline
Type (I)  & $-$388.44 & $-$246.70 \\
Type (II) & $-$439.49 & $-$268.34 \\
\bottomrule
\end{tabular}
\end{center}
\end{table}
\normalsize
In Table~\ref{tab:ads-ene}, the adsorption energies are listed. All the adsorption energies are negative, indicating that all the adsorptions are exothermic reactions. We see that the adsorption on Type (II) surface is slightly more favorable than on Type (I) at both 0.5 ML and 0.25 ML, although the Type (I) surface formation from the bulk needs more energy than the Type (II) surface.

Figure~\ref{fig:ads} shows the optimized atomistic structure of ZOD adsorption complexes on HAP (001) surfaces at 0.25 ML coverage. It is observed that in the Type (I) surface the hydrogen atom of ZOD's phosphonate group moved to oxygen atom of HAP surface's phosphate group, making formation of additional OH group on the surface, and thus indicating that the adsorption is kind of chemisorption caused by proton exchange. There are several hydrogen bonds between ZOD's phosphonate OH group and oxygen atoms of the surface's phosphate groups, and vice versa. It is also found that Ca1 and Ca2 atoms with 6 coordination number (CN) make bonds with oxygen and nitrogen atoms of ZOD, respectively. Such bonding was expected based on the surface relaxation analysis and HOMO-LUMO distribution of isolated ZOD molecule. In the case of Type (II) surface, there are also hydrogen bonds between the surface and ZOD, but the move of hydrogen atom is not observed, which indicates that the adsorption is kind of physisorption. Nevertheless, the magnitude of adsorption energy on Type (II) surface is larger than that on Type (I) surface.

\begin{figure*}[!ht]
\begin{center}
\includegraphics[clip=true,scale=0.3]{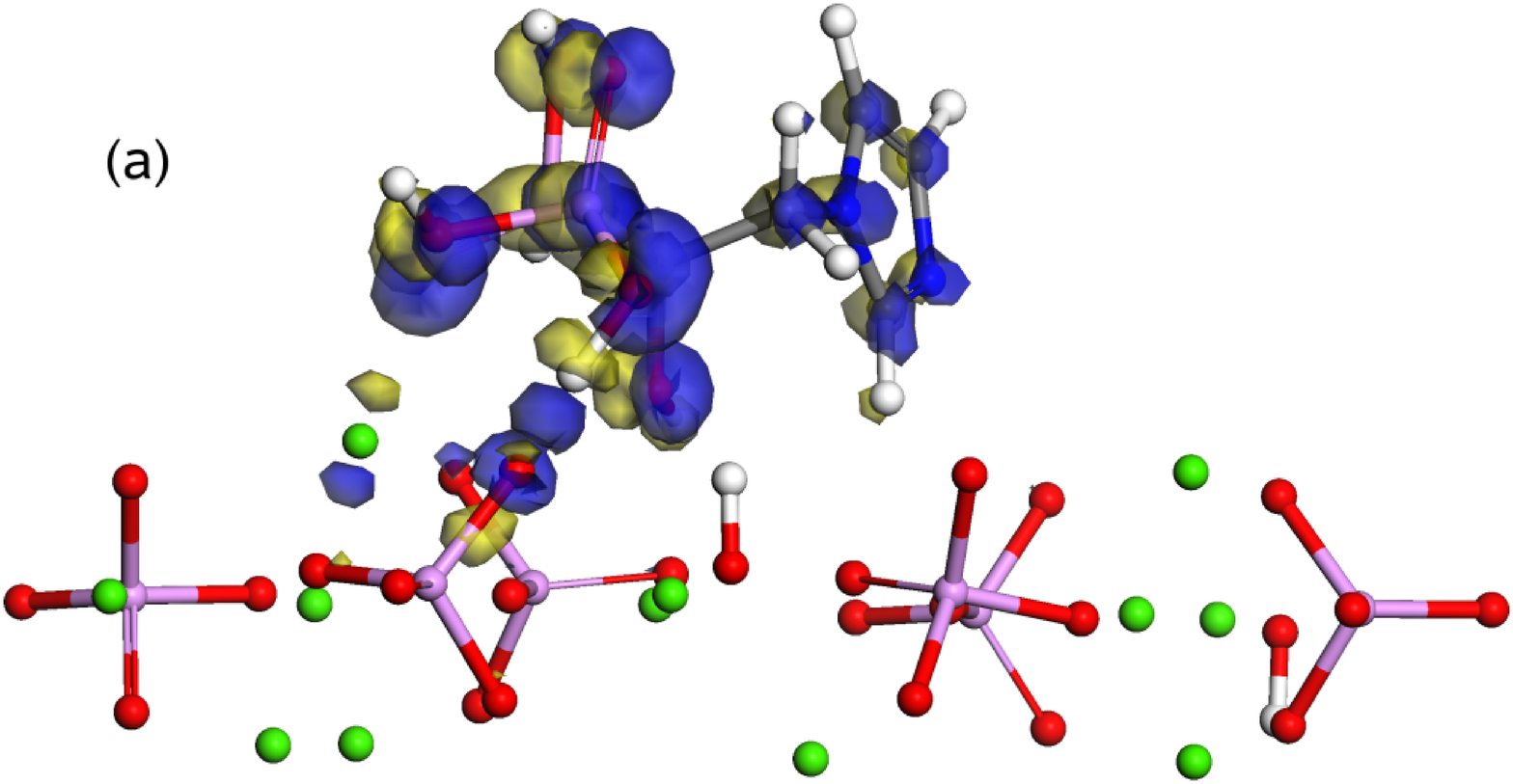} \\
\includegraphics[clip=true,scale=0.23]{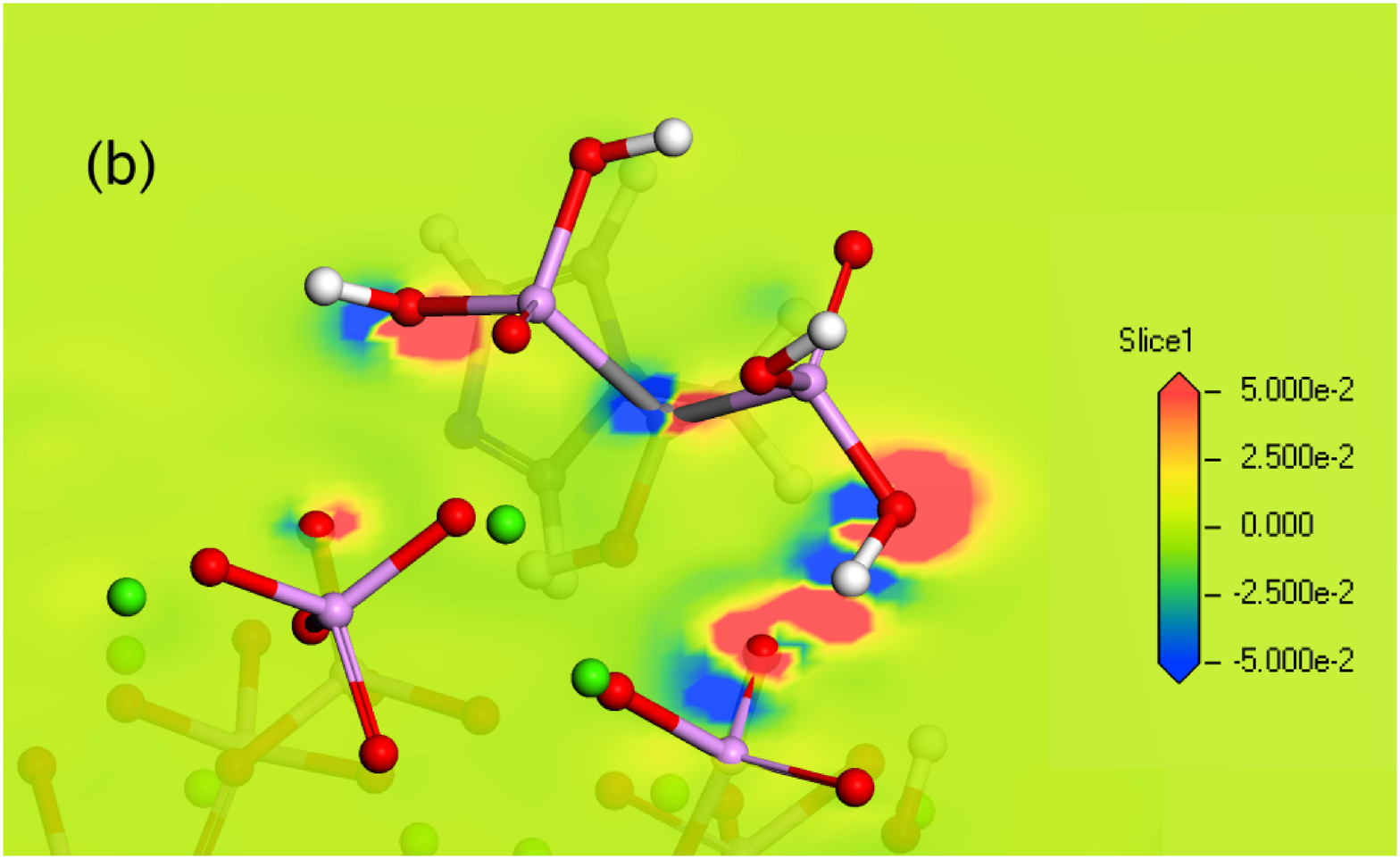}
\includegraphics[clip=true,scale=0.23]{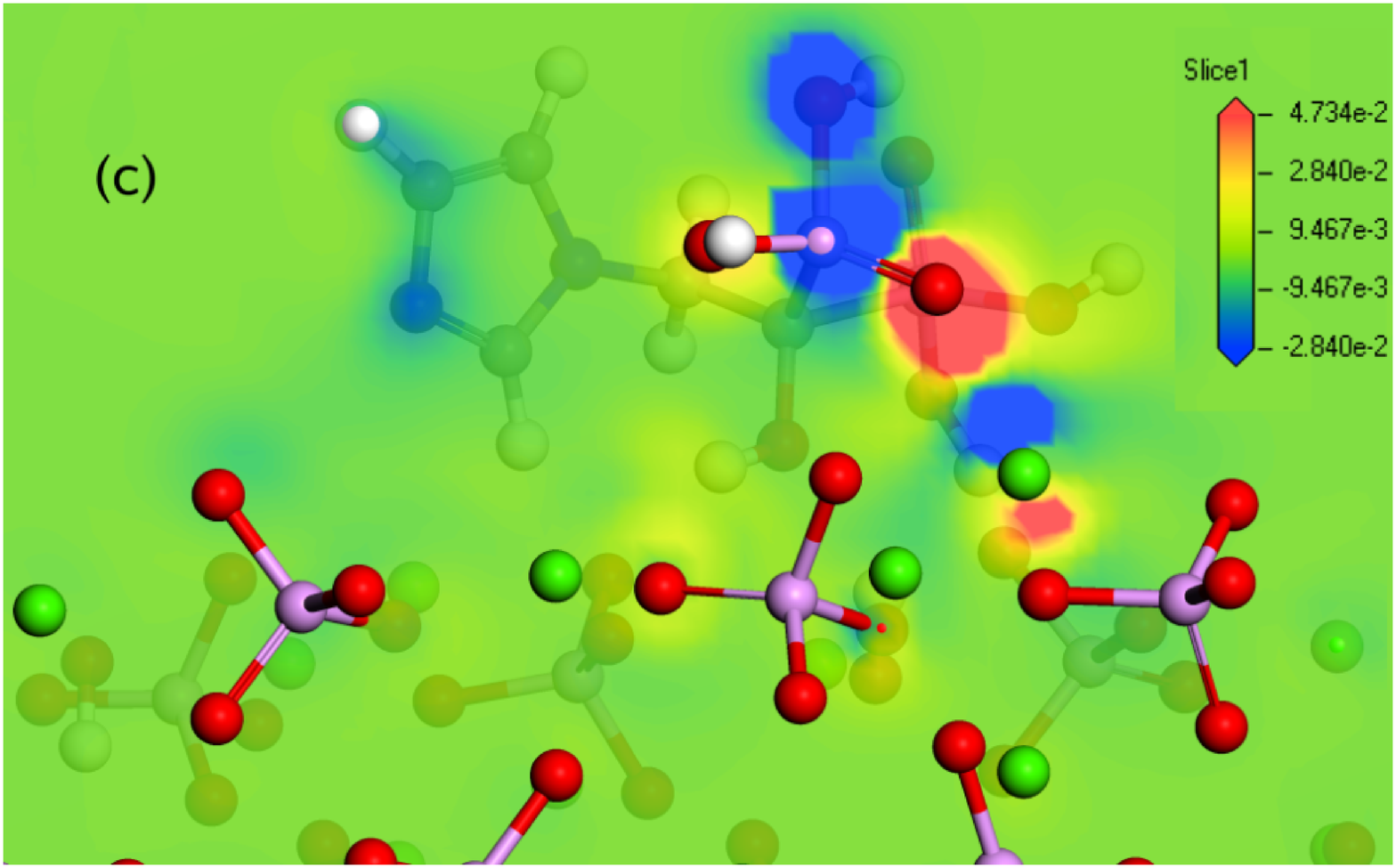}
\caption{\label{fig:diffrho}(Color online) Electronic charge density difference of zoledronic acid adsorption complex on hydroxyapatite (001) surface. (a) isosurface figure with the value of $\pm$0.04 eV/\AA$^3$ (yellow for + and blue for - value), (b) contour figure on the plane containing hydrogen bond between O of phosphonate group of HAP surface and hydrogen of phosphate group, and (c) contour figure around Ca cation. (Ca: green, P: purple, O: red, N: blue, H: white)}
\end{center}
\end{figure*}

To make it clear the charge transfer occurred during the adsorption, we show the electron density difference, $\rho_\text{surf+mol}-(\rho_\text{surf}+\rho_\text{mol})$, plots in Figure~\ref{fig:diffrho}, where (a) is for the isosurface figure with the value of $\pm$0.04 eV/\AA$^3$, and (b) and (c) are the contour plots on the planes around hydrogen bond and Ca cation, respectively. These figures clearly give the evidence of charge transferring at the event of adsorption, from hydrogen of phosphate group of ZOD molecule to oxygen of PH$_4$ group of HAP surface making formation of hydrogen bond, and from Ca of HAP surface to oxygens of phosphonate group forming the coordinate bond. Therefore the adsorption of ZOD on the HAP (100) surface is surely chemisorption.

\begin{figure}[!ht]
\begin{center}
\includegraphics[clip=true,scale=0.55]{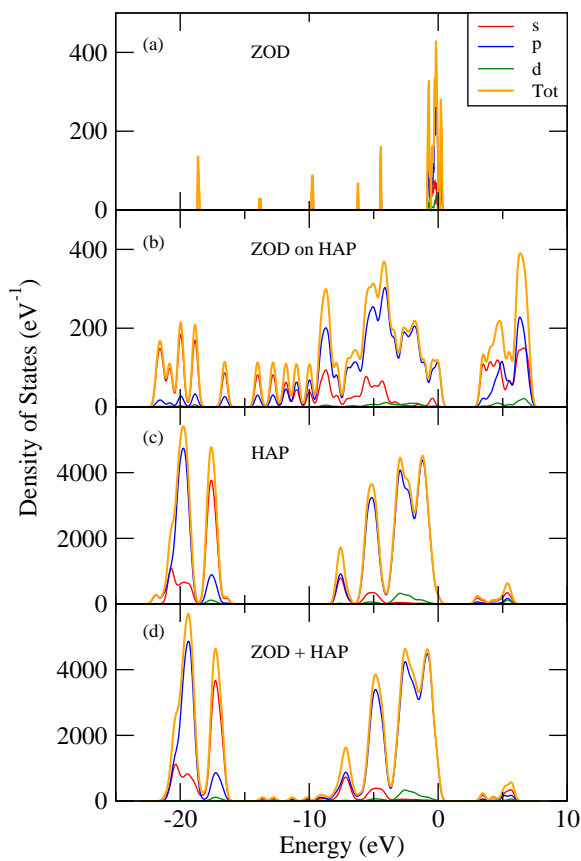}
\caption{\label{fig:dos}(Color online) Total and partial density of states in isolated zoledronic acid (a), adsorbed zoledronic acid (b), hydroxyapatite (100) surface (c) and zoledronic acid adsorption complex on the surface (d).}
\end{center}
\end{figure}

\begin{figure}[!ht]
\begin{center}
\includegraphics[clip=true,scale=0.55]{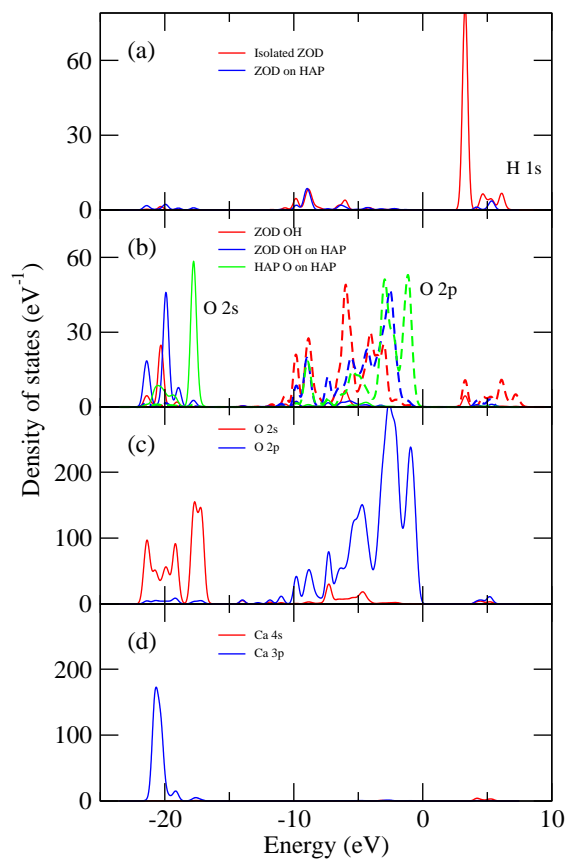}
\caption{\label{fig:lpdos}(Color online) Atom projected partial density of states; (a) for 1s electrons of hydrogen in different situations, (b) for 2s and 2p electrons of oxygen in different situations, (c) for 2s and 2p states of oxygens surrounding Ca cation in ZOD adsorbed HAP surface, and (d) for 4s and 3p of calcium forming the coordinate bond on HAP surface.}
\end{center}
\end{figure}

We calculated the density of states (DOS) of electrons, total and partial DOS shown in Figure~\ref{fig:dos} and atom projected and partial DOS in Figure~\ref{fig:lpdos}. In Figure~\ref{fig:dos} we see the energy spectrum of isolated ZOD molecule (a), the DOS of ZOD molecule that was adsorbed on HAP surface (b), the DOS of HAP (100) surface (c), and the whole complex system comprised of the adsorbed ZOD molecule and HAP (100) surface. Some hybridization between ZOD and HAP surface electrons is shown in the figures. Which atoms cause the charge transferring at the adsorption? To make an answer to this question, we carefully investigate the atom projected partial DOS. The 1s peak of hydrogen of isolated ZOD molecule placed over 0 eV (red line in Figure~\ref{fig:dos} (a) panel) is remarkably weakened after the adorption (blue line in (a) panel), indicating the loss of electrons from hydrogen and the gain by oxygen, and thus the formation of hydrogen bond between hydrogen atom of ZOD and oxygen of HAP surface. We also see the hybridizations between hydrogen 1s state and oxygen 2p state in (a) and (b) panels, and between oxygen 2s state and calcium 3p state in (c) and (d) panels.

\section{\label{sec:con}Conclusion}
In conclusion, we have attempted to make a modeling of adsorption of zoledronic acid on hydroxyapatite (001) surface to get an atomistic insight of bone protection. The systematic study has been performed, from hydroxyapatite bulk and surface, and zoledronic acid to adsorption of the molecule on the (001) surface. We have carried out the structural optimizations and atomic relaxations of the bulk, molecule, surface, and adsorption complexes on the surfaces, and obtained the structural informations and energetics.

Using the three-dimensional periodic supercell model, we determined the stable conformation of the molecule and calculated the molecular orbitals. It was concluded that two phosphonate groups can play a role of electron acceptor in the chemical reaction, while the heterocyclic ring containing nitrogen atom can be a electron donor. After verifying the validity of computational parameters of SIESTA work through the bulk hydroxyapatite calculation, surface modeling and relaxations were performed, and surface formation energies were calculated for two kinds of (001) surface terminations, which are about 1.2 and 1.5 J/m$^2$. Subsequently, the adsorption of zoledronic acid on the relaxed surface was studied, obtaining the surface binding structure and calculating the adsorption energies. We found that the surface Ca atoms and oxygen atoms of phosphate groups can form surface bond including hydrogen bond and coordinate bond with nitrogen, hydrogen, and oxygen atoms of zoledronic acid. The calculated adsorption energies are about -260 kJ/mol at 0.25 ML coverage and -400 kJ/mol at 0.5 ML coverage, indicating the strong binding affinity of zoledronic acid to hydroxyapatite surface.

We have made an interpretation of such strong binding affinity through the analysis of atomistic structures, electron density difference, the density of states and Hirshfeld charges of atoms relevant to surface binding. The results showed that bond lengths and bond angles of phosphonate group of zoledronic acid are similar to those of phosphate group of hydroxyapatite bulk and surface, indicating the strong binding affinity related to the structural similarity. It was also found that the charge transferring is occurred mainly on the side of phosphonate group in one kind surface, while in another surface it is occurred on both imidazol ring containing nitrogen atom and phosphonate group of zoledronic acid.

\section*{\label{ack}Acknowledgments}
The simulations have been carried out on the HP Blade System c7000 (HP BL460c) that is owned and managed by Faculty of Materials Science, Kim Il Sung University. This work was supported from the Commette of Education (grant number 02-2014), DPR Korea.

\section*{\label{note}Notes}
The authors declare that they have no conflict of interest.

\bibliographystyle{elsarticle-num-names}
\bibliography{Reference}

\end{document}